\documentstyle[12pt]{article}
\newcommand\be{\begin{equation}}
\newcommand\ee{\end{equation}}
\newcommand\bea{\begin{eqnarray}}
\newcommand\eea{\end{eqnarray}}
\newcommand\ket[1]{|#1\rangle}

\newcommand\braket[2]{\langle #1|#2\rangle}
\newcommand{\fatalpha}{{\bf \alpha \kern -0.44em \alpha}}
\newcommand{\fatsigma}{{\bf \sigma \kern -0.54em \sigma}}
\newcommand{\tpchi}{{\bf \chi \kern -0.35em \chi}}
\newcommand{\llambda}{{\bf \lambda \kern -0.45em \lambda}}

\renewcommand{\theequation}{\arabic{equation}}
\renewcommand{\theequation}{\thesection-\arabic{equation}}
\bibliography{plain}
\title{\bf Continuous-time quantum walks on
star graphs}\vspace{20mm}
\author{ S. Salimi
  \thanks{Corresponding author:  E-mail addresses: shsalimi@uok.ac.ir}
 \\ {\small Department of Physics,
University of Kurdistan, P.O.Box 66177-15175 , Sanandaj, Iran.}}
\pagebreak

\vspace{20mm}
\begin{document}
\maketitle \vspace{15mm}
\newpage
\begin{abstract}
 In this paper, we investigate continuous-time
quantum walk on star graphs. It is shown that quantum central limit
theorem for a continuous-time quantum walk on star graphs for
$N$-fold star power graph, which are invariant under
 the quantum component of adjacency matrix, converges to continuous-time quantum
walk on  $K_2$ graphs (Complete graph with two vertices) and the
probability of observing walk  tends to the uniform distribution.


{\bf PACs Index: 03.65.Ud }
\end{abstract}

\vspace{70mm}
\newpage
\section{Introduction}
Quantum walks were introduced in the early 1990s by Aharonovich,
davidovich and Zaggury \cite{adz}. Since then the topic has
attracted considerable  interest. The continuing attraction can be
traced back to at least two reasons. First, the quantum walk is of
sufficient interest in its own right because there are fundamental
differences compared to the classical random walk. Next, quantum
walks offer quite a number of possible applications. One of the best
known is the link between quantum walks and quantum search
algorithms which are superior to their classical
counterparts\cite{ccdfgs, Kempe}. Similar to classical random walk
there are two types of quantum walks, discrete and continuous time
\cite{adz,Kempe1}. A study of quantum walks on simple graph is well
known in physics (for more details see \cite{fls}). Recent studies
of quantum walks on more general graphs were described in
\cite{ccdfgs, fg,cfg,abnvw,aakv,mr,kem,khovi}. Some of these works
studies the problem in the important context of algorithmic problems
on graphs and suggests that quantum walks is a promising algorithmic
technique for designing future quantum algorithms. One approach for
investigation of continuous-time quantum walk (CTQW) on graphs is
using the spectral distribution associated with the adjacency matrix
of graphs \cite{js,jsa, jsas, konno1, konno2, salimi1,salimi2}.
Authors in Refs.\cite{js,jsa} have introduced a new method for
calculating the probability amplitudes of quantum walk based on
spectral distribution. In this method a canonical relation between
the Fock space  of stratification graph and set of orthogonal
polynomials has been established where leads to obtain the
probability measure (spectral distribution) adjacency matrix graph.
 The method of spectral distribution
only requires simple structural data of graph and allows us to avoid
a heavy combinational argument often necessary to obtain full
description of spectrum of the adjacency matrix.

In this paper we investigate CTQW on star graphs, which are
discussed by Burioni et, al. \cite{Burioni1,Burioni2} as models of
the Bose-Einstein condensation, by quantum probability theory. To
this aim, first by turning the graph into metric space based on
distance function, we have been able to stratify and quantum
decomposition \cite{js,nob}, such that the basis of Hilbert space of
quantum walk consist of superposition of quantum kets of vertices
belonging to the same stratum with the same coefficient (Fock
space). Then  by applying an isomorphism from Fock space onto the
closed linear span of the orthogonal polynomials with respect to the
measure $\mu$ we obtain spectral distribution adjacency matrix graph
and probability amplitudes of observing walk at strata. Also we try
to investigate quantum central limit theorem for CTQW in $N$-fold
star power graphs as $N\longrightarrow\infty$, and show the CTQW
converges to CTQW on $K_2$ (Complete graph with two vertices) where
the probability of observing walk  tends to the uniform
distribution. On the other hand, quantum walker induces to the graph
that spectral distribution it tend to the Bernoulli distribution \
$\mu\longrightarrow\frac{1}{2}(\delta_1+\delta_{-1})$.

The organization of the paper is as follows: we give a brief review
of graph, star graph, stratification and quantum decomposition in
the section $2$. Section $3$ is devoted to study CTQW for some
example of star graphs via quantum probability theorey and try to
investigate quantum central limit theorem for CTQW on their graphs.
In the conclusion we summarize the obtained results and discuss
possible development. Finally, in the appendix the determination of
spectral distribution associated with adjacency matrix  by Stieltjes
transform is derived.

\section{Star graph, Stratification and Quantum decomposition}
Let $V$ be a non-empty set and $E$ be a subset of $\{\{\alpha,
\beta\}| \alpha, \beta\in V \ \mbox{for}\ \alpha\neq \beta \}$. The
pair $G=(V,E)$ is called a graph, where elements of $V$ and $E$ are
 vertices and edges of graph, respectively. We say that two
vertices of $\alpha$ and $\beta$ are adjacent if $\{\alpha,
\beta\}\in E$ and write $\alpha\sim \beta$. A finite sequence
$\alpha_0, \alpha_1,..., \alpha_n$ is said a walk of length $n$ if
$\alpha_k\sim \alpha_{k+1}$ for $k=0, 1, ..., n-1$. A graph is
called connected if any pair of distinct vertices is connected by a
walk. The degree or valency of a vertex $\alpha\in V$ is defined by
\begin{equation}
\kappa(\alpha)=|\{\beta \in V|\ \beta\sim \alpha \}|
\end{equation}
where $|.|$ denote the cardinality. For a graph $G$  the adjacency
matrix $A$ is given by
\[
A_{\alpha \beta} = \left\{
\begin{array}{ll}
1 & \mbox{if $ \alpha\sim \beta$}\\
0 & \mbox{otherwise.}
\end{array}
\right.
\]

Obviously, (i) $A$ is a symmetric (ii)  elements of $A$  take a
value in $\{0, 1\}$ (iii) diagonal elements of $A$ are $0$.
Conversely, for a non-empty set $V$, a structure graph is uniquely
determined by a such matrix which indexed by $V$. On the other hand,
$A$ is considered as an operator acting on the Hilbert space
$l^2(V)$ in such a way that
$$
A\ket{\alpha}=\sum_{\alpha\sim \beta}\ket{\beta}, \;\;\;\;\alpha\in
V,
$$
where $\{\ket{\alpha}|\ \alpha \in V\}$ forms a complete orthogonal
basis of $l^2(V)$.

In this paper, we focus on  investigating CTQW on star graphs which
are also discussed by Burioni et al. \cite{Burioni1,Burioni2} as
models of the Bose- Einstein condensation. A star graph is obtained
from star product of graphs which we define in the following.

Let us consider two graphs $G^\nu,\ \nu=1,2$ with adjacency matrices
$A^\nu$, each of which is equipped with an origin $o_\nu\in
V^{(\nu)}$. Define a matrix $A$ as
\begin{equation}
A_{(\alpha,\beta),(\acute{\alpha},\acute{\beta})}=A^{(1)}_{\alpha
\acute{\alpha}}\delta_{\beta o_2}\delta_{\acute{\beta}
o_2}+\delta_{\alpha o_1}\delta_{\acute{\alpha} o_1}A^{(2)}_{\beta
\acute{\beta}}, \quad \alpha,\acute{\alpha}\in V^{(1)}, \quad \beta,
\acute{\beta}\in V^{(2)}
\end{equation}
We observe that $A$ is symmetric, elements  take a value in $\{0,
1\}$ and the diagonal consists of $0$. Therefore $A$ is the
adjacency matrix of a graph. The connected component containing
$(o_1, o_2)$ is called the star product of $G^{(1)}$ and  $G^{(2)}$
 and the resulted graph is a star graph \cite{nob}(see Fig.1).
 One can generalize this definition for $\nu=1,2,...,N$ that we
 consider $N$-fold star power graphs for our aims. The
stratification is introduced \cite{js,nob} by taking
$o=(o_1,o_2,...,o_N)$ as the origin and have
\begin{equation}\label{v1}
V=\bigcup_{k=0}^{\infty}V_k,\;\;\;\;\;\; V_i=\{\alpha\in V|\
\partial(o,\alpha)=k\}.
\end{equation}
Here $\partial(\alpha, \beta)$ stands for the length of the shortest
walk connecting $\alpha$ and $\beta$. According to the
stratification (\ref{v1}), we define a unit vector by
\begin{equation}
\ket{\phi_{k}}=\frac{1}{\sqrt{|V_k|}}\sum_{\alpha\in V_{k}}\ket{k,
\alpha},
\end{equation}
where $\ket{k, \alpha}$ denotes the eigenket of the $\alpha$-th
vertex at the stratum $k$ and let $\Gamma(G)$ the closed subspace of
$l^2(V)$ spanned by $\{\ket{\phi_{k}}\}$. Moreover, the
stratification (\ref{v1})give rise to a quantum decomposition:
\begin{equation}\label{qd1}
A=A^{+}+A^{-}+A^0.
\end{equation}
If $\Gamma(G)$ is invariant under the actions of the quantum
components quantum components $A^\varepsilon$, $\varepsilon\in
\{+,-,0\}$,  then there exist two Szeg\"{o}- Jacobi sequences
$\{\omega_k\}_{k=1}^{\infty}$ and $\{\alpha_k\}_{k=1}^{\infty}$
derived from $A$, such that
\begin{equation}\label{v5}
A^{+}\ket{\phi_{k}}=\sqrt{\omega_{k+1}}\ket{\phi_{k+1}}, \;\;\;\
k\geq 0
\end{equation}
\begin{equation}\label{v6}
A^{-}\ket{\phi_{0}}=0, \;\;\
A^{-}\ket{\phi_{k}}=\sqrt{\omega_{k}}\ket{\phi_{k-1}}, \;\;\;\ k\geq
1
\end{equation}
\begin{equation}\label{v7}
A^{0}\ket{\phi_{k}}=\alpha_{k+1}\ket{\phi_{k}}, \;\;\;\ k\geq 0.
\end{equation}
The above coefficients (i.e., Szeg\"{o}- Jacobi sequences ) are
obtained from geometric feature of graph \cite{js,nob}. Then
$(\Gamma(G), A^+, A^-, A^o )$ is an interacting Fock space
associated with Szeg\"{o}- Jacobi sequences $\{\omega_k,
\alpha_k\}$.
\section{CTQW on star graph via quantum probability theory }
In the study of CTQW on graphs, the spectral distribution of $A$ in
the vacuum state $\ket{\phi_0}$ plays an essential role \cite{js},
which is by definition a probability distribution $\mu$ uniquely
specified by
\begin{equation}
\langle A^m\rangle= \langle \phi_0| A^m|\phi_0\rangle=\int x^m
\mu(dx), \quad m=0,1,2,...,
\end{equation}
where, according to \cite{js, jsa,nob}, $\langle A^m\rangle$
coincides with the number of $m$-step walks starting and terminating
at origin point $o$. For analyzing the spectral distribution $\mu$
of adjacency matrix $A$, we use the method of quantum decomposition
which is a powerful tool. The spectral distribution $\mu$ is
determined by applying the canonical isomorphism from the
interacting Fock space onto the closed linear span of orthogonal
polynomials determined by  Szeg\"{o}- Jacobi sequences $\{\omega_k,
\alpha_k\}$. In fact the determination of $\mu$ is the main problem
in the spectral theory of operators, where in the case of star
graphs  is quite possible by using the Stieltjes method, as it is
explained in appendix A. Then by using the quantum decomposition
relations (\ref{qd1})-(\ref{v7}) and the recursion relation of
polynomials $P_n(n)$(\ref{op}), the other matrix elements as
\begin{equation}\label{prob1}
\langle \phi_k|
A^m|\phi_0\rangle=\frac{1}{\sqrt{\omega_1\omega_2...\omega_k}}\int
x^m P_k(x)\mu(dx), \quad m=0,1,2,....
\end{equation}
One of our main goals in this paper is the evaluation of probability
amplitudes of CTQW by using Eq.(\ref{prob1}), such that we have
\begin{equation}\label{prob2}
q_k(t)=\langle \phi_k|
e^{-itA}|\phi_0\rangle=\frac{1}{\sqrt{\omega_1\omega_2...\omega_k}}\int
e^{-itx} P_k(x)\mu(dx),
\end{equation}
where $|q_{k}(t)|^2$ is the probability of observing the walk at the
stratum $k$ at time $t$. The conservation of probability
$\sum_{k}|q_{k}(t)|^2=1$ follows immediately from Eq. (\ref{prob2})
by using the completeness relation of orthogonal polynomials
$P_n(x)$. In the appendix $A$ reference \cite{js} is  provided the
walker has the same probability  at the all sites belonging to the
same stratum, i.e., we have $|q_{ik}(t)|^2=\frac{|q_k(t)|^2}{|V_k|},
\ \mbox{for}\ i\in V_k$, where $|q_{ik}(t)|^2$ denotes the
probability of the walker at the $i$-th vertex of $k$-th stratum
$V_k$. Investigation of CTQW via spectral distribution method, which
is introduced as a new method for calculating the probability
amplitudes quantum walk (for more details see \cite{js} ), allows us
to avoid a heavy combinational argument often necessary to obtain
full description of spectrum of the Hamiltonian.

We can now investigate CTQW on star graphs. In the first, we
calculate CTQW on two simple star graph  and study quantum central
limit theorem for them. In the end, we study this question for star
lattice.

\subsection{Examples 1.}
In this example we investigate the CTQW on  the star finite path
graph with three vertices as Fig.2. Then we have

\begin{equation}
\omega_1=N, \quad\omega_2=1; \quad \alpha_n=0
\end{equation}
and
\begin{equation}
G_{\mu_N}(z)=\frac{z^2-1}{z^3-(N+1)z}, \quad
\mu_N=\frac{1}{N+1}\delta_0+\frac{N}{2(N+1)}\left(\delta_{\sqrt{N+1}}+\delta_{-\sqrt{N+1}}\right).
\end{equation}
Therefore the amplitudes probability of CTQW on this graph are as
follows:
$$
q_0(t)=\int e^{-ixt}\mu(dx)=\frac{1}{N+1}[1+N\cos(\sqrt{N+1}t)]
$$
$$
q_1(t)=\frac{1}{\sqrt{N}}\int x
e^{-ixt}\mu(dx)=-i\sqrt{\frac{N}{N+1}}\sin(\sqrt{N+1}t)
$$
\begin{equation}
q_2(t)=\frac{1}{\sqrt{N}}\int (x^2-N)
e^{-ixt}\mu(dx)=\frac{N}{(N+1)\sqrt{N}}[-1+\cos(\sqrt{N+1}t)].
\end{equation}

{\bf{Quantum central limit theorem}}

Having studied CTQW on this graph for $N$ arbitrary, we investigate
CTQW   in the limit of large $N\longrightarrow\infty$ which in fact
we discuss  this question as a  quantum central limit theorem for
CTQW. To state a  quantum central limit theorem for CTQW we
have\cite{konno1}
\begin{equation}
q_k(t)=\lim_{N\longrightarrow\infty}\braket{\phi_k}{e^{\frac{-itA}{\sqrt{N}}}|\phi_0}=\lim_{N\longrightarrow\infty}\frac{1}{\sqrt{\omega_1\omega_2...\omega_k}}\int
P_k(x) e^{-ixt/\sqrt{N}}\mu_N(x)dx
\end{equation}

Then we obtain the amplitudes probability as
$$
q_0(t)=\lim_{N\longrightarrow\infty}\int_R
e^{\frac{-itx}{\sqrt{N}}}\mu(dx)=
\lim_{N\longrightarrow\infty}\int_R
e^{\frac{-itx}{\sqrt{N}}}(\frac{1}{N+1}\delta_0+\frac{N}{2(N+1)}(\delta_{\sqrt{N+1}}+\delta_{-\sqrt{N+1}}))dx
$$
\begin{equation}
=\frac{1}{2}(e^{-it}+e^{it})= \cos t,
\end{equation}
$$
q_1(t)=\lim_{N\longrightarrow\infty}\frac{1}{\sqrt{N}}\int_R
xe^{\frac{-itx}{\sqrt{N}}}\mu(dx)=\lim_{N\longrightarrow\infty}\frac{1}{\sqrt{N}}\int_R
xe^{\frac{-itx}{\sqrt{N}}}(\frac{1}{N+1}\delta_0+\frac{N}{2(N+1)}(\delta_{\sqrt{N+1}}+\delta_{-\sqrt{N+1}}))dx
$$
\begin{equation}
\frac{1}{2}(e^{-it}-e^{it})= -i\sin t,
\end{equation}
\begin{equation}
q_2(t)=0,
\end{equation}
where the probability of the observing walk ($|q_{i}(t)|^2$, for
$i=0,1$ ) reaches the uniform distribution periodically at $t=k\pi +
\frac{\pi}{4}$ for $k\in Z$.

Indeed, in the limit of large $N\longrightarrow\infty$(e.i., quantum
central limit theorem ), the CTQW on this graph is reduced to CTQW
on $K_2$ (Complete graph with two vertices). In other words, the
spectral distribution $\mu_N$ associated with the adjacency matrix
of the graph tend to the Bernoulli distribution
($\mu_N\longrightarrow\frac{1}{2}(\delta_1+\delta_{-1})$), where is
the spectral distribution of $K_2$ graphs.

\subsection{Examples 2.}
In the second example we consider CTQW on the star graph as Fig.3.
Then we have
\begin{equation}
\omega_1=2N, \quad\omega_2=2; \quad \alpha_n=0
\end{equation}
and
\begin{equation}
G_{\mu_N}(z)= {\frac {{x}^{2}-2}{x \left( {x}^{2}-2(N+1) \right)
}},\quad
\mu_N=\frac{1}{N+1}\delta_0+\frac{N}{2(N+1)}\left(\delta_{\sqrt{2(N+1)}}+\delta_{-\sqrt{2(N+1)}}\right).
\end{equation}
The amplitudes probability of CTQW on this graph are as:
$$
q_0(t)=\int e^{-ixt}\mu(dx)=\frac{1}{N+1}[1+N\cos(\sqrt{2(N+1)}t)]
$$
$$
q_1(t)=\frac{1}{\sqrt{2N}}\int x
e^{-ixt}\mu(dx)=-i\sqrt{\frac{N}{N+1}}\sin(\sqrt{2(N+1)}t)
$$
\begin{equation}
q_2(t)=\frac{1}{2\sqrt{N}}\int (x^2-N)
e^{-ixt}\mu(dx)=\frac{N}{2(N+1)\sqrt{N}}[-1+2\cos(\sqrt{N+1}t)].
\end{equation}
Also in this example,  quantum central limit theorem for CTQW is
reduced on CTQW on complete graph $K_2$.
\subsection{Examples 3.}
{\bf{Star Lattice}}

The star lattice is the $N$-fold star power of finite path graph,
see Fig.4. One can show that the two sequence $\{\omega_k\}$ and
$\{\alpha_k\}$ obtain as
\begin{equation}
\omega_1=N, \quad\omega_2=\omega_3=\cdots=1; \quad \alpha_k=0
\end{equation}
Substituting sequence $\omega_k$ and $\alpha_k$ in (5.22), the
Stieltjes transform $G_{\mu_N}(z)$ of spectral distribution $\mu_N$
takes the following form
\begin{equation}\label{q1}
G_{\mu_N}(z)=
\frac{1}{z-\frac{N}{z-\frac{1}{z-\frac{1}{z-\frac{1}{\ddots}}}}}
\end{equation}
In order to evaluate the continued fraction, we need first to
evaluate the following infinite continued fraction defined as
\begin{equation}
\tilde{G}_{\mu}(z)=
\frac{1}{z-\frac{1}{z-\frac{1}{z-\frac{1}{z-\frac{1}{\ddots}}}}}=\frac{1}{z-\tilde{G}_{\mu}(z)},
\end{equation}
where by solving above equation we have
\begin{equation}\label{q2}
\tilde{G}_{\mu}(z)=\frac{1}{2}(z-\sqrt{z^2-4}).
\end{equation}
Then by substituting (\ref{q2}) in (\ref{q1}), we obtain the
following expression for the Stieltjes transform of $\mu$
\begin{equation}\label{q1}
G_{\mu_N}(z)=
\frac{1}{z-N\tilde{G}_{\mu}(z)}=\frac{1}{2}\frac{(2-N)z-N\sqrt{z^2-4}}{N^2-(N-1)z^2}.
\end{equation}
Finally by applying Stieltjes inversion formula, we acquire the
absolutely continuous part of spectral distribution $\mu_N$ as
follows
\begin{equation}\label{q1}
\mu_N(x)=\frac{1}{2\pi}\frac{N\sqrt{4-x^2}}{N^2-(N-1)x^2} \ ; \qquad
-2\leq x\leq 2.
\end{equation}
Now we investigate CTQW on some of the known infinite graphs can be
obtained from star lattice  by appropriate choice of $N$ as:

{\bf{A.}} For $N=1$ we obtain finite path graph. Then we have
$$
\omega_1=1, \quad\omega_2=\omega_3=\cdots=1; \quad \alpha_k=0
$$
\begin{equation}
\mu_1(x)=\frac{1}{2\pi}\sqrt{4-x^2}\ ; \qquad -2\leq x\leq 2,
\end{equation}
where the amplitudes probability for CTQW on this graph are
\begin{equation}
q_k(t)=i^k(J_k(t)+ J_{k+2}(t))=2i^k(k+1)\frac{J_{k+1}(t)}{t};\quad
k= 0, 1, . . .
\end{equation}
(for more detail see Ref.\cite{js}).

{\bf{B.}} For $N=2$ we obtain infinite line graph where the two
sequences, spectral distribution  and the amplitude probability for
observing walk at strata at time $t$ are
$$
\omega_1=2, \quad\omega_2=\omega_3=\cdots=1; \quad \alpha_k=0
$$
$$
\mu_2(x)=\frac{1}{\pi}\frac{1}{\sqrt{4-x^2}}\ ; \qquad -2\leq x\leq
2,
$$
\begin{equation}
q_0(t)=J_0(t);\qquad q_k(t)=i^k\sqrt{2}J_k(t), \quad k\geq 1,
\end{equation}
(for more detail see Ref.\cite{js}).

{\bf{C.}} For $N\geq  3$ the total mass of $\mu_N(x)$ is less than
one (i.e., $\int \mu_N(x)dx\leq 1$). In  fact, it contains a
discrete measure where by a simple computation one can obtain as
\cite{nob1}
\begin{equation}\label{q1}
\mu_N(x)=\frac{1}{2\pi}\frac{N\sqrt{4-x^2}}{N^2-(N-1)x^2}+\frac{N-2}{2N-2}(\delta_{\frac{N}{\sqrt{N-1}}}+\delta_{\frac{-N}{\sqrt{N-1}}})
\ ; \qquad -2\leq x\leq 2.
\end{equation}
{\bf{Quantum central limit theorem}}

Now we consider behavior CTQW on this graph for large $N$ (i.e.,
$N\longrightarrow\infty$) where we discuss this question as a
quantum central limit theorem for CTQW.

The quantum central limit theorem for CTQW is
$$
q_{k}(t)=\lim_{N\longrightarrow\infty}\braket{\phi_k}{e^{-iAt/\sqrt{N}}|\phi_0}=
\lim_{N\longrightarrow\infty}\frac{1}{\sqrt{N}}\int P_k(x)
e^{-ixt/\sqrt{N}}\mu_N(x)dx
$$
$$
=\lim_{N\longrightarrow\infty}\frac{1}{\sqrt{N}}\int P_k(x)
e^{-ixt/\sqrt{N}}\left(\frac{1}{2\pi}\frac{N\sqrt{4-x^2}}{N^2-(N-1)x^2}+\frac{N-2}{2N-2}(\delta_{\frac{N}{\sqrt{N-1}}}+
\delta_{\frac{-N}{\sqrt{N-1}}})\right)dx
$$
$$
=\lim_{N\longrightarrow\infty}\frac{1}{\sqrt{N}}\int P_k(x)
e^{-ixt/\sqrt{N}}\frac{N-2}{2N-2}(\delta_{\frac{N}{\sqrt{N-1}}}+
\delta_{\frac{-N}{\sqrt{N-1}}})dx
$$
\begin{equation}
=\lim_{N\longrightarrow\infty}\frac{N-2}{2N-2}\left(
\frac{P_k(\frac{N}{\sqrt{N-1}})}{\sqrt{N}}e^{\frac{-itN}{\sqrt{N(N-1)}}}+
\frac{P_k(\frac{-N}{\sqrt{N-1}})}{\sqrt{N}}e^{\frac{itN}{\sqrt{N(N-1)}}}
\right).
\end{equation}
By a simple computation one can show that
$\lim_{N\longrightarrow\infty} \frac{P_k(\frac{\pm
N}{\sqrt{N-1}})}{\sqrt{N}}=0$ for $k\geq 2$. Then we have
$$
q_0(t)=\frac{1}{2}(e^{-it}+e^{it})= \cos t,
$$
$$
q_1(t)=\frac{1}{2}(e^{-it}-e^{it})= -i\sin t,
$$
\begin{equation}
q_k(t)=0 \quad \mbox{for} \quad k\geq 2.
\end{equation}
We see that quantum central limit theorem for CTQW on the star
lattice is reduced on complete graph $K_2$ and the amplitudes
probability reach  the uniform distribution. Really, for every star
graph that is invariant under the quantum component $A^\epsilon,\
\epsilon\in \{+,-,o\}$, one can show that quantum central limit
theorem for CTQW on it is reduced CTQW on complete graph $K_2$.

\section{Conclusion}
In this work, we have investigated CTQW on star graphs, which are
discussed by Burioni et, al. \cite{Burioni1,Burioni2} as models of
the Bose-Einstein condensation, by quantum probability theory.
First, we have equipped graph to metric based on distance function.
Then we have stratified  graph and according to stratification
accomplished quantum decomposition of adjacency matrix. The basis of
Hilbert space of quantum walk form a Fock space where by applying an
isomorphism from it onto the closed linear span of the orthogonal
polynomials  we have obtained spectral distribution adjacency matrix
graph and probability amplitudes of observing walk at strata. Also,
we have studied quantum central limit theorem for CTQW on star
graphs and shown the CTQW converges to CTQW on $K_2$ (complete graph
with two vertices), such that the probability of observing walk tend
to the uniform distribution.  Indeed, we can generalize CTQW on star
graphs which are invariant under
 the quantum component $A^\epsilon,\ \epsilon\in \{+,-,o\}$, quantum central limit
theorem for them induce in  CTQW on graphs that the spectral
distribution it tend to the Bernoulli distribution.

\vspace{1cm} \setcounter{section}{0}
 \setcounter{equation}{0}
 \renewcommand{\theequation}{A-\arabic{equation}}
  {\Large{Appendix A}}\\
\textbf{\large{Determination of spectral distribution by the
Stieltjes transform }}

In this appendix we explain how we can determine spectral
distribution $\mu(x)$ of the graphs, by using the Szeg\"{o}-Jacobi
sequences $(\{\omega_k\},\{\alpha_k\})$. To this aim we may apply
the canonical isomorphism from the interacting Fock space onto the
closed linear span of the orthogonal polynomials determined by the
Szeg\"{o}-Jacobi sequences $(\{\omega_i\},\{\alpha_i\})$. More
precisely, the spectral distribution $\mu$ under question is
characterized by the property of orthogonalizing the polynomials
$\{P_n\}$ defined recurrently by
$$ P_0(x)=1, \;\;\;\;\;\
P_1(x)=x-\alpha_1,$$
\begin{equation}\label{op}
xP_n(x)=P_{n+1}(x)+\alpha_{n+1}P_n(x)+\omega_nP_{n-1}(x),
\end{equation}
for $n\geq 1$.

As it is shown in \cite{tsc}, the spectral distribution ì can be
determined by the following identity:
\begin{equation}\label{v3}
G_{\mu}(z)=\int_{R}\frac{\mu(dx)}{z-x}=\frac{1}{z-\alpha_1-\frac{\omega_1}{z-\alpha_2-\frac{\omega_2}
{z-\alpha_3-\frac{\omega_3}{z-\alpha_4-\cdots}}}}=\frac{Q_{n-1}^{(1)}(z)}{P_{n}(z)}=\sum_{l=1}^{n}
\frac{A_l}{z-x_l},
\end{equation}
where $G_{\mu}(z)$ is called the Stieltjes transform and $A_l$ is
the coefficient in the Gauss quadrature formula corresponding to the
roots $x_l$ of polynomial $P_{n}(x)$ and where the polynomials
$\{Q_{n}^{(1)}\}$ are defined
recurrently as\\
        $Q_{0}^{(1)}(x)=1$,\\
    $Q_{1}^{(1)}(x)=x-\alpha_2$,\\
    $xQ_{n}^{(1)}(x)=Q_{n+1}^{(1)}(x)+\alpha_{n+2}Q_{n}^{(1)}(x)+\omega_{n+1}Q_{n-1}^{(1)}(x)$,\\
    for $n\geq 1$.

Now if $G_{\mu}(z)$ is known, then the spectral distribution $\mu$
can be recovered from $G_{\mu}(z)$ by means of the Stieltjes
inversion formula:
\begin{equation}\label{m1}
\mu(y)-\mu(x)=-\frac{1}{\pi}\lim_{v\longrightarrow
0^+}\int_{x}^{y}Im\{G_{\mu}(u+iv)\}du.
\end{equation}
Substituting the right hand side of (\ref{v3}) in (\ref{m1}), the
spectral distribution can be determined in terms of $x_l,
l=1,2,...$, the roots of the polynomial $P_n(x)$, and  Guass
quadrature constant $A_l, l=1,2,... $ as
\begin{equation}\label{m}
\mu=\sum_l A_l\delta(x-x_l)
\end{equation}
 ( for more details see Refs. \cite{js,jsa,tsc,st}.)

\newpage
{\bf Figure Captions}

{\bf Figure-1:} Star graph.

{\bf Figure-2:} Figure.2

{\bf Figure-3:} Figure.3.

{\bf Figure-4:} Star lattice for $N=8$

\end{document}